\documentclass[pra ,showpacs,showkeys,twocolumn]{revtex4-1}

\usepackage{graphicx}
\usepackage{latexsym}
\usepackage{amsmath}
\usepackage{amssymb}
\usepackage{amsfonts}
\usepackage{color}
\usepackage[normalem]{ulem}

\begin{document}

%\preprint{}

\title{Multivortex states and dynamics in nonequilibrium quantum fluids}

\author{Vladimir N. Gladilin}

\author{Michiel Wouters}

\affiliation{TQC, Universiteit Antwerpen, Universiteitsplein 1,
B-2610 Antwerpen, Belgium}

\date{\today}

\begin{abstract}
{In strongly nonequilibrium Bose-Einstein condensates described by
the generalized Gross-Pitaevskii equation, vortex motion becomes
self-accelerated while the long-range vortex-antivortex interaction
appears to be repulsive. We numerically study the impact of these
rather unusual vortex properties on the dynamics of multivortex
systems. We show that at strong nonequilibrium the repulsion between
vortices and antivortices leads to a dramatic slowdown of their
annihilation. Moreover, in finite-size samples, relaxation of
multivortex systems can lead to the formation of metastable
vortex-antivortex clusters, whose shape and size depend, in
particular, on the sample geometry, boundary conditions and
deviations from equilibrium. We further demonstrate that at strong
nonequilibrium the interaction of self-accelerated vortices with
inhomogeneous condensate flows can lead to generation of new
vortex-antivortex pairs.}
\end{abstract}


 \maketitle


\section{Introduction}

{Topological defects, ubiquitous in ordered phases of matter,
ranging from superfluids an superconductors to neutron stars and
even cosmology, have been investigated for many decades
\cite{chaikin}. Especially noteworthy discoveries have been the
Abrikosov vortex lattice, the Berezinskii-Kosterlitz-Thouless (BKT)
phase transition in two-dimensional superfluids and the Kibble-Zurek
(KZ) mechanism. An Abrikosov vortex lattice is formed when a
superfluid is subject to rotation or a superconductor is put in a
magnetic field. A large number of vortices is then created and they
arrange in a regular lattice because of their repulsive
interactions. The BKT mechanism is driven by thermally activated
vortex antivortex pair dissociation, that results in a transition
from a superfluid state with quasi-long range order to a normal
state. In the KZ mechanism on the other hand, topological defects
are created by a rapid quench through the phase transition. They
subsequently annihilate on a much longer time scale.}

{In by far most of the systems that have been investigated, the
assumption of (local) thermal equilibrium can be assumed to hold. A
natural extension of these studies is then the consideration of
ordered systems that are subject to continuous driving and
dissipation. An experimental motivation for this question stems from
the study of exciton-polariton condensates \cite{carusotto-ciuti}.
In order to compensate for the finite polariton lifetime,
excitations have to be continuously injected. This driving and
dissipation prevents the system from reaching local equilibrium. At
the same time, an out-of-equilibrium system does not imply an
unstable system since stability is obtained through the balance of
dissipation and gain.}

{Spurred by these experimental developments
\cite{Kasprzak06,lagoudakis08,lagoudakis2011,Liew15,caputo2016},
several theoretical works have analyzed the properties of vortices
polariton condensates \cite{marchettiPRL10,cancellieriPRB14,ma2017}.
A more recent research activity has been devoted to the
understanding of the BKT transition out of equilibrium. In one of
these works \cite{dagdavorj}, it was numerically found that the
usual scenario survives the introduction of driving and dissipation
in a finite size system, where the renormalisation group based
studies \cite{wachtel, sieberer} found modifications at large length
scales. More recently, it was shown how the different regimes can be
reached in a polariton quantum fluid in the parametric oscillation
regime \cite{zamora}. The modification of the transition can be
related to a change in the interaction between vortices and
antivortices, that can become repulsive at large distances for weak
nonequilibrium \cite{wachtel} to entirely repulsive at strong
driving and dissipation \cite{30}. This profound change in
interaction between vortices of opposite sign can be attributed to
radial current flows that are emitted from the vortex cores in a
nonequilibrium condensate. Moreover, for stronger nonequilibrium,
the vortex motion becomes self-accelerated: a vortex is dragged by
its own slowly relaxing flow field. }

{The rather unusual properties of individual vortices and pairwise
vortex interactions, revealed in~\cite{30} can be expected not only
to affect the BKT transition, but can also lead to the formation of
new vortex lattice structures. The existence of such stable
complexes could strongly affect the phase ordering kinetics
\cite{kulczy} after a KZ quench \cite{michalKZ} through the phase
transition. Specifically, we will address the following questions in
this paper. How are the dynamics of multivortex systems modified as
a result of the vortex-antivortex repulsion? Can this repulsion lead
to the appearance of metastable vortex-antivortex configurations in
pinning-free nonequilibrium condensates, and what are those
configurations? What happens with self-accelerated vortices when the
deviation from equilibrium further increases? In particular, can a
self-accelerated vortex, similarly to a fast moving obstacle,
produce new vortex-antivortex pairs, akin to a von K\'{a}rm\'{a}n
street~\cite{24,25,26})? }

The paper is organized as follows. In Sec. II we describe the
theoretical model and the numerical scheme used in our calculations.
The processes of vortex-antivortex recombination in nonequilibrium
polariton condensates are considered in Sec. III. Section IV deals
with the formation of various metastable vortex-antivortex states in
finite-size samples. In Sec. V we analyze the effects produced by
fast moving vortices in strongly nonequilibrium polariton
superfluids. The main conclusions from the obtained results are
summarized in Sec. VI.

\section{Model}

We describe the nonresonantly excited two-dimensional polariton
condensate by the generalized Gross-Pitaevskii equation that is
obtained after adiabatic elimination of the exciton reservoir
\cite{15,16,a31}
\begin{eqnarray}
({\rm i}-\kappa)\hbar \frac{\partial \psi}{\partial t} =&&
\left[-\frac{\hbar^2\nabla^2}{2m} +g |\psi|^2 \right. \nonumber
\\
&&\left.+\frac{{\rm i}}{2} \left(\frac{P}{1+|\psi|^2/n_s}-\gamma
\right) \right] \psi, \label{ggpe}
\end{eqnarray}
Here $m$ is the effective mass and the contact interaction between
polaritons is characterized by the strength $g$. The imaginary term
in the square brackets on the right hand side describes the
saturable pumping (with strength $P$ and saturation density $n_s$)
that compensates for the losses ($\gamma$). {We take into account
that the energy relaxation $\kappa$ in the condensate~\cite{38,39},
may be non-negligible. It originates from scattering with thermal
phonons and excitons and results in the appearance of damping in the
vortex dynamics.} In particular, this damping tends to impede vortex
drag by a condensate flow so that the vortex core velocity becomes
smaller than the velocity of the surrounding condensate. This
results in the appearance of a non-zero Magnus force, exerted on a
vortex and perpendicular to the flow direction. Thus, for two
vortices of the same (opposite) chirality, moving in each other
velocity field, the Magnus forces lead to mutual repulsion
(attraction).

{With the help of the Madelung transformation $\psi=\sqrt{n}e^{i
\theta}$, a continuity equation can be derived for the polariton
density $n$
\begin{equation} (1+\kappa^2)\frac{\partial n}{\partial
t}+\mathbf{\nabla}\cdot(n \mathbf{v}) = (R -2 \kappa \epsilon) n.
\end{equation}
Here, we introduced the velocity
$\mathbf{v}=\mathbf{\nabla} \theta /m$, the net gain
$R=P/(1+n/n_s)-\gamma$ and the single particle energy
$\epsilon=gn+\frac{m}{2}\mathbf{v}^2+\frac{\nabla^2\sqrt{n}}{2m\sqrt{n}}$.
The right hand side of the continuity equation shows that the net
gain acts as a source of particles. The increase of $R$ with
decreasing density implies that a particle source is present where
the density is locally suppressed, such as at a vortex core or at
the sample boundary. }

In general, the pumping intensity $P$ is coordinate dependent and
can be represented as $P({\bf r})=P_0p({\bf r})$, with
$P_0=\max{P({\bf r})}$. Since the interaction strength is positive,
$g>0$, it is convenient to rewrite Eq.~(\ref{ggpe}) in a
dimensionless form, by expressing the particle density $|\psi|^2$ in
units of $n_0\equiv n_s(P_0/\gamma-1 )$, time in units of
$\hbar/(gn_0)$, and length in units of $\hbar/\sqrt{2mgn_0}$:
\begin{eqnarray}
({\rm i}-\kappa)\frac{\partial\psi}{\partial t}=&& \left[-\nabla^2
+|\psi|^2 + \phantom{\frac{\psi^2}{\psi^2}}\right. \nonumber
\\
&&\left.+{\rm i}c\frac{1-|\psi|^2+(1+\nu^{-1})(p-1)}{1+\nu |\psi|^2}
\right] \psi . \label{ggpe2}
\end{eqnarray}
Equation~(\ref{ggpe2}) contains three dimensionless scalar
parameters: $\kappa$ characterizes, as described above, damping in
the system dynamics, $c=\gamma/(2gn_s)$ is a measure of the
deviation from equilibrium, and $\nu=n_0/n_s$ is proportional to the
relative excess of the maximum pumping intensity $P_0$ over the
threshold intensity. The dimensionless function $p({\bf r})$ ($0\leq
p \leq 1$) describes the spatial distribution of the pumping
intensity.

Equation (\ref{ggpe2}) is solved numerically using a
finite-difference scheme. One of the key elements of our
approach~\cite{Silhanek11} is the use of automatically adapted time
step $h_t$. The adaptation is aimed at minimizing the number of
steps in $t$ and - at the same time - at keeping the solving
procedure accurate. This allows us to combine an accurate and
detailed description of fast stages in the system dynamics with a
high time efficiency in the case when the dynamics is relatively
slow. The proposed approach has been successfully applied to
describe different kinds of vortex dynamics in superconducting
condensates (see, e. g., \cite{23,40,29}) as well as the motion of
vortex-(anti)vortex pairs and individual vortices in nonequilibrium
polariton condensates~\cite{30}. Here we use a uniform
two-dimensional grid with the step $h=0.2$ to 0.5. The time step
$h_t$ is typically $\sim 10^{-5}$ to $10^{-3}$ depending on the
specific distribution of the order parameter.

Our simulations are performed for infinite periodic systems and
finite-size samples. In the latter case we use the Dirichlet
boundary conditions $\psi|_{\rm boundary}=0$. Due to suppression of
the polariton density at the boundary, radiative polariton losses
are locally reduced. For this reason, at uniform pumping the
boundary becomes a source of extra polaritons, which propagate
inside the sample (cp. radial flows emitted by a vortex
core~\cite{30}). These inward currents tend to drag vortices away
from the boundaries, providing an effective confinement for
vortices. The strength of this confinement can be tuned by changing
the pump intensity near the boundary. In the present calculations we
consider the intensity distribution in the form of
\begin{equation}
p(x,y)=1-\alpha {\rm e}^{-d(x,y)/d_0}, \label{pump}
\end{equation}
where $d(x,y)$ is the distance to the nearest sample boundary and
the positive parameter $\alpha$ is smaller than 1. With increasing
$\alpha$ and/or $d_0$ the inward currents weaken and can even turn
to outward flows at sufficiently large $\alpha$. {The latter case,
where vortices are removed very fast from the sample, impeding the
study of vortex dynamics, is not studied in the present work. }

\section{Vortex-antivortex annihilation}

It is natural to expect that the vortex-antivortex repulsion, caused
by the outflow of condensate particles from a vortex core, can
significantly affect the processes of vortex-antivortex
recombination in non-equilibrium condensates \cite{wachtel,30}.
{Vortices in non-equilibrium condensates can {emerge} in different
schemes. Apart from the spontaneous generation of vortices induced
by noise \cite{dagdavorj}, vortices can appear due to resonant
excitation with a Gauss-Laguerre beam~\cite{tosi11,marchetti10},
creation of polariton condensates using chiral lenses~\cite{dall14},
or by a rapid quench of the pump power, in analogy to the
Kibble-Zurek mechanism \cite{michalKZ,Comaron17,lagoudakis11}.}

\begin{figure} \centering
\includegraphics*[width=0.9\linewidth]{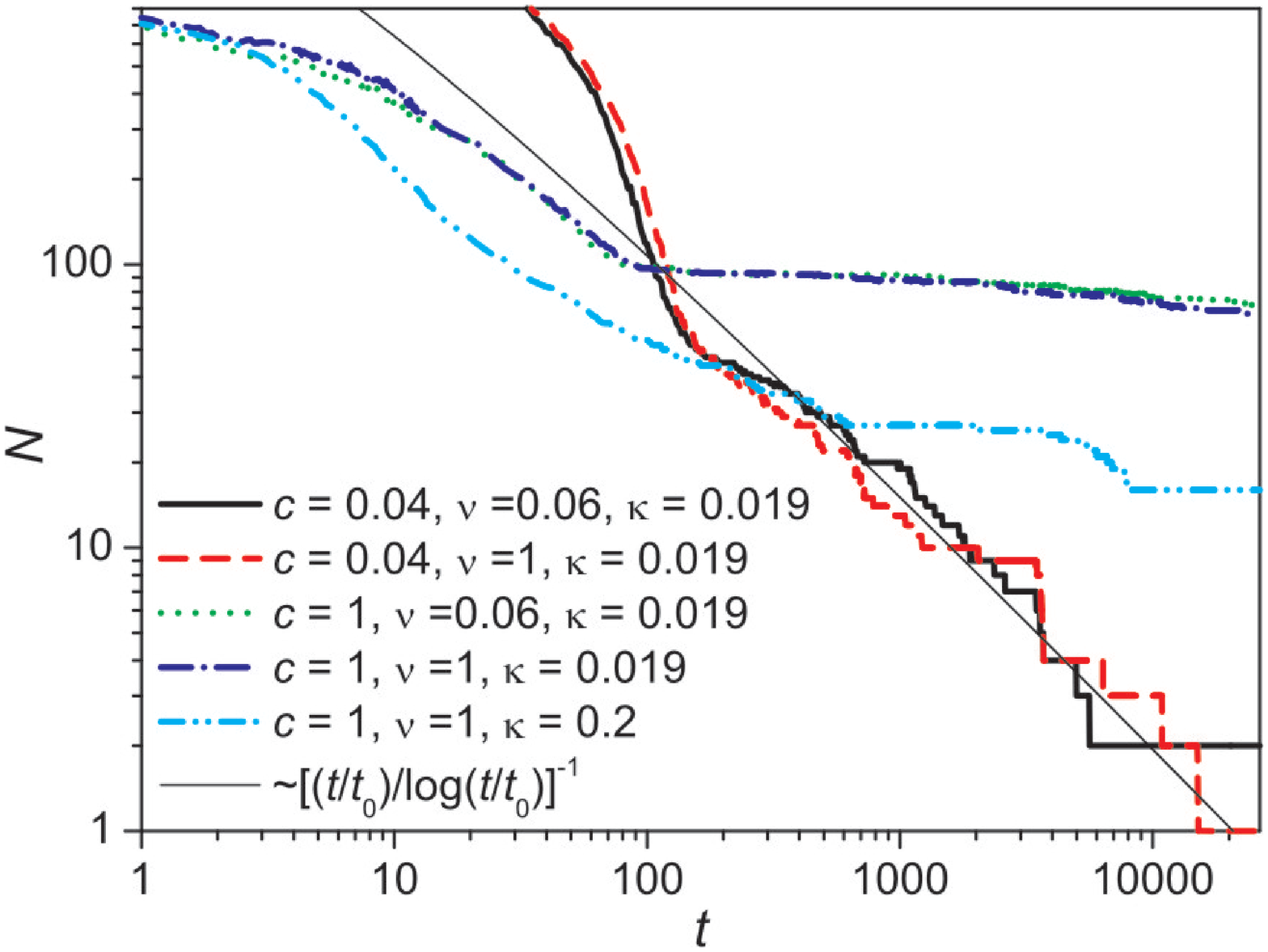}
\caption{Number of vortex pairs per unit cell as a function of time
in a periodic system with period $L_x=L_y=165$ at different values
of $c$, $\nu$, and $\kappa$ (thick lines). The calculations are
performed with the uniform pumping $p\equiv 1$ and grid step
$h=0.55$. The thin line corresponds to a decay $\sim
[(t/t_0)/\log(t/t_0)]^{-1}$ with $t_0=0.33$. \label{relaxation1}}
\end{figure}

{After the injection of a certain number of vortices, the subsequent
phase healing dynamics typically consists of the annihilation of
vortex-antivortex pairs, leading to a decay of the number of
vortices \cite{kulczy,Comaron17}. } In Fig.~\ref{relaxation1} we
show the calculated time dependence of the number of vortex pairs
per unit cell of a periodic system with period $L_x=L_y=165$. The
initial conditions correspond to the presence of 1000 vortices and
1000 antivortices, randomly distributed within the unit cell. At
$c=0.04$, $\nu=0.06$, and $\kappa=0.019$ our system practically
coincides with one of those analyzed in~\cite{Comaron17}. Quite in
line with the results of~\cite{Comaron17}, for this set of
parameters the simulated late-time dynamics closely follows the
dependence $N\sim [(t/t_0)/\log(t/t_0)]^{-1}$ with $t_0=0.33$. This
dynamic does not change too much when the parameter $\nu$
considerably increases (see the dashed curve in
Fig.~\ref{relaxation1}).

{On the other hand}, an increase of the nonequilibrium parameter to
$c=1$ results in a dramatic slowing down of the vortex-antivortex
annihilation at relatively large $t$ (see the dotted curve for
$c=1$, $\nu=0.06$, and $\kappa=0.019$). Like in the case of weak
non-equilibrium ($c=0.04$), the behavior of $N(t)$ at $c=1$ is
rather insensitive to the value of $\nu$ (compare the dotted and
dash-dotted curves to each other). At the same time, the
annihilation processes can be somewhat accelerated by increasing the
damping parameter $\kappa$ (see the dash-dot-dot curve in
Fig.~\ref{relaxation1}) As described above, stronger damping
enhances the Magnus effect and the corresponding attractive
component in the vortex-antivortex interaction. Nevertheless, even
for $\kappa$ as large as 0.2, the average annihilation rate at $c=1$
is seen to be by orders of magnitude lower than that in the nearly
equilibrium system with $c=0.04$.

\begin{figure} \centering
\includegraphics*[width=0.9\linewidth]{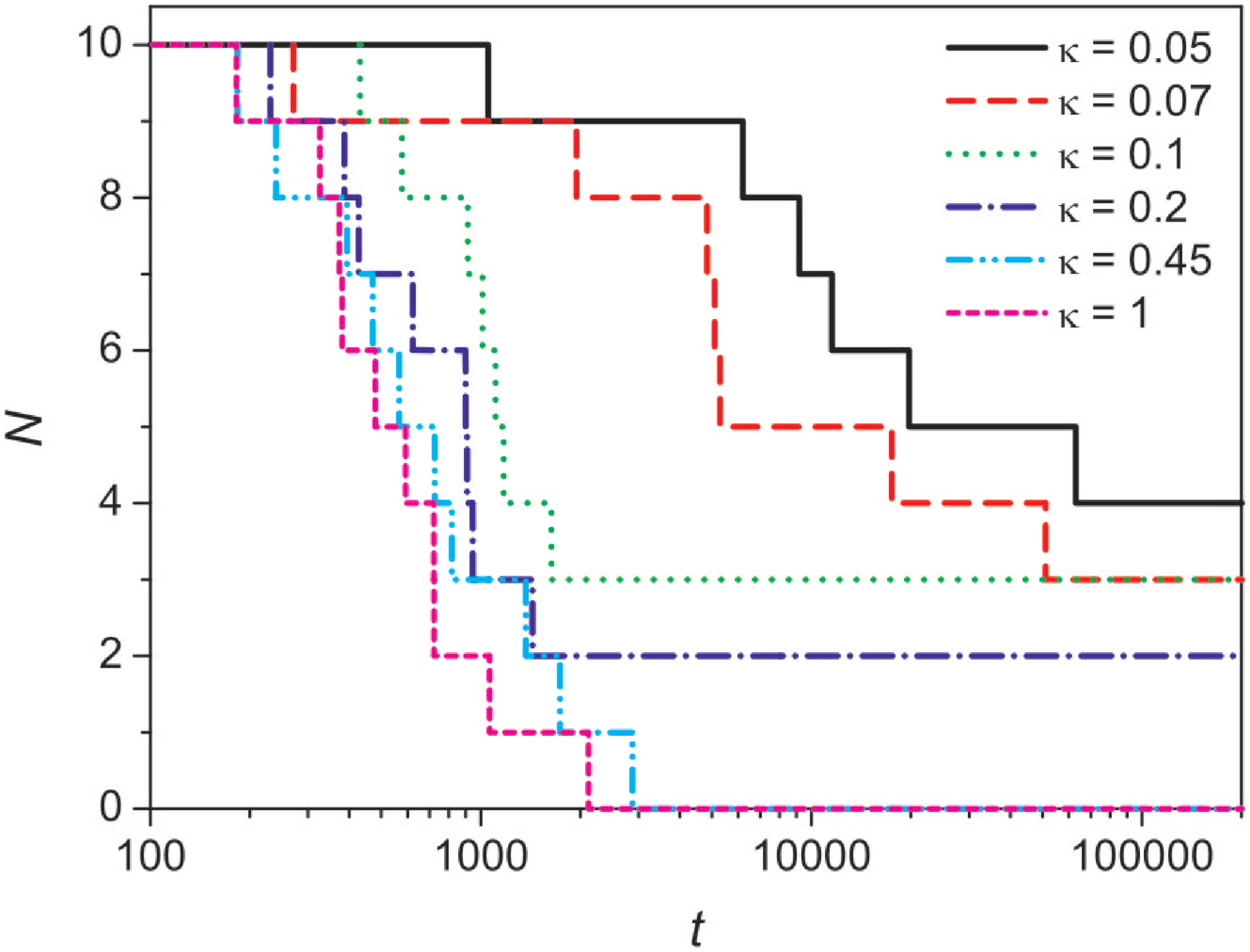}
\caption{Number of vortex pairs as a function of time in a square
with sides $L_x=L_y=150$ at $\nu=1$, $c=0.3$ and different values of
$\kappa$. The calculations are performed with the Dirichlet boundary
conditions, uniform pumping ($p\equiv 1$) and grid step $h=0.5$.
\label{relaxation2}}
\end{figure}

{Our simulations show that the repulsion between vortices and
antivortices can considerably slow down their recombination already
at moderate deviations from equilibrium. Figure~\ref{relaxation2}
illustrates this effect for the more experimentally relevant case of
a finite-size sample. In this figure we show the time evolution of
the number $N$ of vortex pairs in a square-shaped sample at $c=0.3$}
and different values of the damping parameter $\kappa$. All the
curves correspond to one and the same quasi-random initial
distribution of 10 pinned vortices and 10 pinned antivortices. Like
in~\cite{30}, the pinning potential for a vortex is provided by an
ad hoc requirement $ \psi = 0$ on a single point of the numerical
grid, starting from an initial condition with unit circulation
(clockwise or counterclockwise). At the time moment $t=0$ this
pinning potential is removed so that vortex motion and annihilation
processes become possible.

{As seen from Fig.~\ref{relaxation2}, a decrease of the damping
parameter $\kappa$ leads to a decrease of the average annihilation
rate for vortices and antivortices that is especially pronounced at
$\kappa<0.1$. This behavior reflects a competition between the
vortex-antivortex repulsion, caused by the outflow of condensate
particles from a vortex core, and the attraction due to Magnus
forces, which are quite strong at $\kappa \sim 1$ but become small
at $\kappa<0.1$.}

{More interestingly, Fig.~\ref{relaxation2} implies that} at
relatively small values of $\kappa$ the evolution of the system does
not lead directly to a vortex-free state. Instead, metastable
vortex-antivortex states are formed, where the number of vortex
pairs tends to increase with decreasing $\kappa$. In the next
section we consider those states in more detail.

\section{Metastable vortex-antivortex states}

\begin{figure} \centering
\includegraphics*[width=0.8\linewidth]{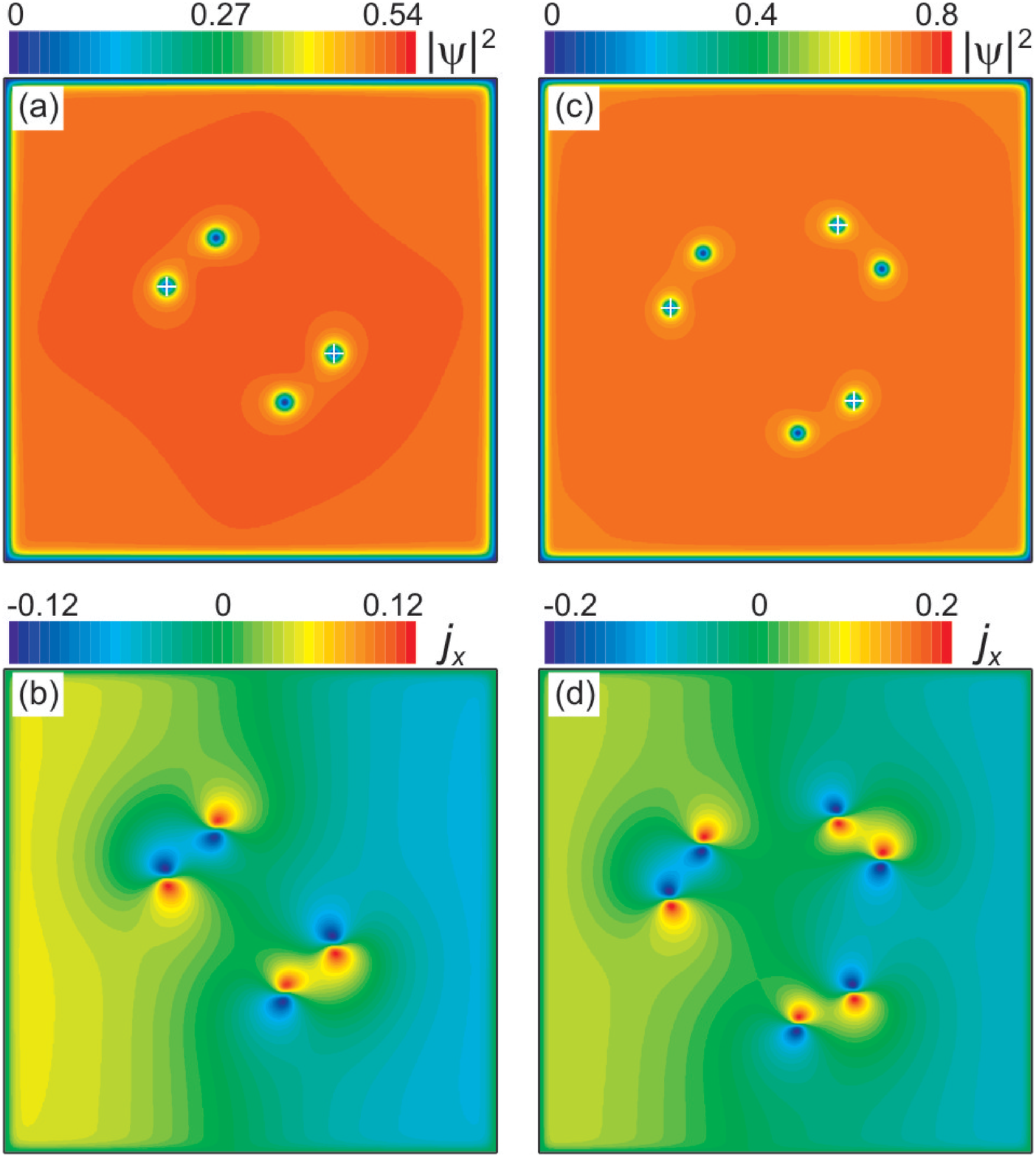}
\caption{Distributions of the particle density [panels (a) and (c)]
and the $x$ component of the current density [panels (b) and(d)]
corresponding to the static metastable configurations of vortex
pairs formed in a square with $L_x=L_y=150$, $\nu=1$ and $c=0.3$ at
$\kappa=0.2$ [panels (a) and (b)] and $\kappa=0.07$ [panels (c) and
(d)]. The calculations are performed with the Dirichlet boundary
conditions, uniform pumping ($p\equiv 1$) and grid step $h=0.5$. The
white crosses in panels (a) and (c) indicate vortices with
counterclockwise circulating currents. \label{static}}
\end{figure}

{Two examples of metastable vortex-antivortex patterns are shown in
Fig. \ref{static}. Panels (a) and (b) correspond to the dash-dotted
line in Fig. \ref{relaxation2} ($\kappa=0.2$), where two
vortex-antivortex pairs remain at long times; panels (c) and (d)
correspond to the dashed line ($\kappa=0.07$) with three remaining
pairs. The upper panels show the condensate density and in the lower
panels the $x$ component of the current density ${\bf j}= {\rm
Im}(\psi^*\nabla \psi)$ is plotted.} The patterns of $j_x(x,y)$
allow one to easily distinguish between clockwise and
counterclockwise vortices. The geometry of the shown vortex
configurations is mainly determined by an interplay of the
vortex-(anti)vortex repulsive interactions and the repulsion of
vortices of both polarities by the sample boundaries. Since the
vortex-vortex repulsion is stronger than that for a vortex and
antivortex~\cite{30}, minimization of the energy requires, first of
all, maximum possible distances between vortices of the same
chirality. The optimal distance between neighboring vortices of
opposite chirality appears smaller, because a reduction of the
distance between a vortex and an antivortex leads to a more
efficient (though, of course, partial) mutual cancelation of their
circulating currents. As a result, the obtained metastable
configurations often look to be built of preformed vortex-anivortex
pairs.

\begin{figure} \centering
\includegraphics*[width=0.8\linewidth]{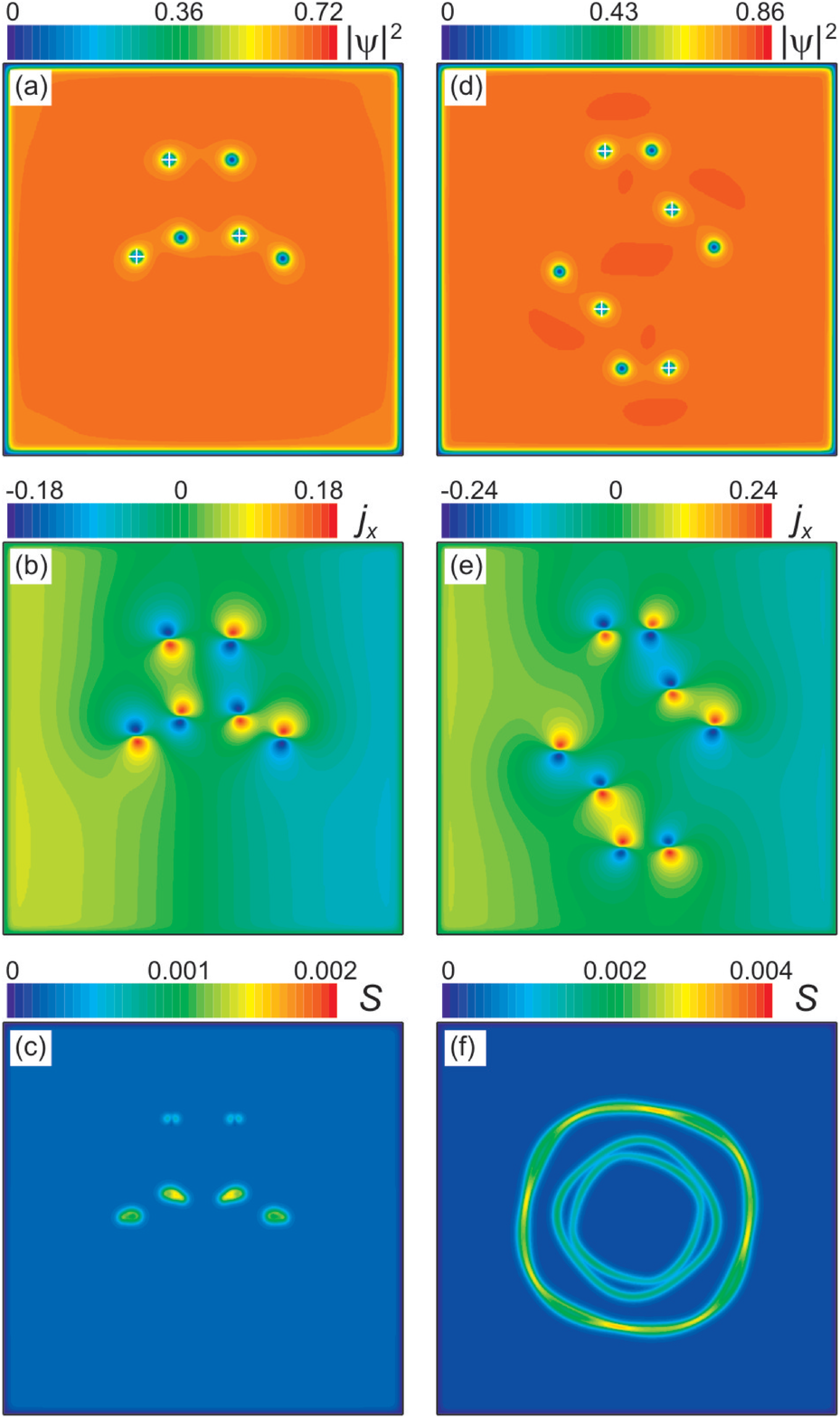}
\caption{Oscillating [panels (a) to (c)] and rotating [panels (d) to
(f)] metastable configurations of vortex pairs formed in a square
with $L_x=L_y=150$, $\nu=1$ and $c=0.3$ at $\kappa=0.1$ and
$\kappa=0.05$, respectively. The plots show snapshots of the
particle-density [panels (a) and (d)] and current-density [panels
(b) and (e)] distributions as well as the distributions of the
parameter $S$ with $t_2-t_1=6000$ [panel (c)] and $t_2-t_1=20000$
[panel (f)]. The calculations are performed with the Dirichlet
boundary conditions, uniform pumping ($p\equiv 1$) and grid step
$h=0.5$. The white crosses in panels (a) and (d) indicate vortices
with counterclockwise circulating currents. \label{oscillating}}
\end{figure}

While the highly symmetric metastable patterns shown in
Fig.~\ref{static} are fully static, at $\kappa=0.1$ and 0.05 the
evolution of the multivortex system, considered above, leads to the
formation of less symmetric oscillating or rotating
vortex-antivortex configurations. This is illustrated by
Fig.~\ref{oscillating}, where we show, together with with snapshots
of the particle-density and current-density distributions, also the
plots of the parameter
\begin{equation}
S(x,y)=\sqrt{\frac{1}{t_2-t_1} \int_{t_1}^{t_2} {\rm d}t \left(
\frac{\partial \vert \psi(x,y,t)\vert^2}{\partial t} \right)^2},
\label{trace}
\end{equation}
which allows to visualize trajectories of moving
vortices~\cite{a41}. For static vortex configurations the values of
$S$ vanish. The $S$-distribution displayed in
Fig.~\ref{oscillating}(c) corresponds to vibrational motion in a
``vortex molecule'' formed by 3 vortex pairs
[Figs.~\ref{oscillating}(a) and (b)]. Rotation of a 4-pair vortex
configuration [Figs.~\ref{oscillating}(d) and (e)] is reflected in
Fig.~\ref{oscillating}(f). As implied by the shape of the vortex
trajectories, this rotation is accompanied by some
confinement-induced deformations of the ``vortex molecule''.

Our simulations show that formation of various metastable
vortex-antivortex states is possible for a rather wide range of the
sample parameters. Few examples of the obtained static vortex
configurations are given below. In all the cases, the simulations
start with a relatively large (40 to 100) number of pairs of
vortices and antivortices, pinned at random positions within the
sample. After removing the pinning potentials, the subsequent
relaxation of the system is simulated until a metastable state is
reached and its stability is confirmed (in particular, by the
corresponding behavior of the parameter $S$).

\begin{figure} \centering
\includegraphics*[width=1.0\linewidth]{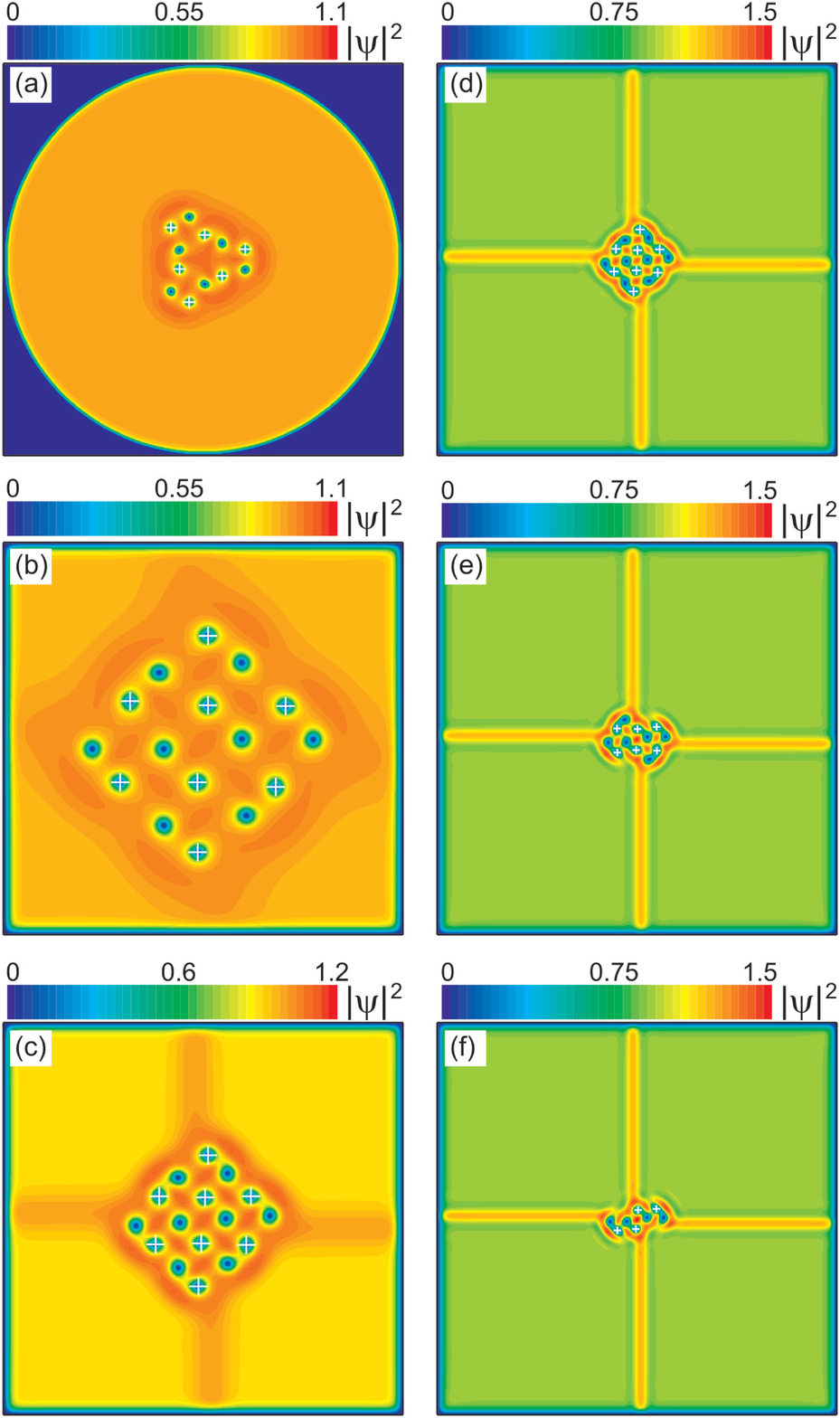}
\caption{(a) Cluster of 6 vortex pairs formed under uniform pumping
($p\equiv 1$) in a disk of radius $R=100$ at  $\nu=1$, $c=1$ and
$\kappa=0.05$. (b)-(f) Clusters of vortex pairs in a square with
sides $L_x=L_y=100$ at $\nu=1$, $\kappa=0.05$ and different values
of $c$: 0.8 (b), 1.3 (c), 3.1 (d), 3.2 (e), and 3.6 (f). The pumping
intensity is described by Eq.~(\ref{pump}) with $\alpha=0.25$ and
$d_0=0.6$. All the calculations are performed with the Dirichlet
boundary conditions and grid step $h=0.5$. The white crosses
indicate vortices with counterclockwise circulating currents.
\label{clusters1}}
\end{figure}

Figure~\ref{clusters1}(a) demonstrates a metastable
vortex-antivortex configuration, formed in a disk with a highly
nonequilibrium polariton condensate ($c=1$). At such a large value
of the non-equilibrium parameter, the inward currents from the
boundaries of the sample strongly push vortices to the center of the
disk. This results in the formation of a rather compact vortex
cluster, consisting of a relatively small number of vortex pairs.

{It is obvious that the distribution of the confining inward
currents and hence the sample geometry can strongly affect the shape
of metastable clusters. This is illustrated by
Fig.~\ref{clusters1}(b), where we show a ``diamond-like'' cluster
stabilized as a result of vortex relaxation in a square with sizes
$L_x=L_y=100$. Due to the use of non-uniform pumping with a reduced
intensity at the sample edges [$\alpha=0.25$ and $d_0=0.6$ in
Eq.~(\ref{pump})], the compressive effect of the inward currents is
somewhat weakened so that the formed cluster is relatively sparse
despite of a rather large value of the nonequilibrium parameter
($c=0.8$)}. The highly symmetric diamond-like shape of the vortex
cluster, shown in Fig.~\ref{clusters1}(b), {persists under a
relatively strong} increase of the confining currents from the
sample boundaries. {In the simulations, reflected in
Fig.~\ref{clusters1}, such an increase has been realized by
increasing $c$ with the step $\delta c=0.1$, starting from the
configuration displayed in Fig.~\ref{clusters1}(b), at fixed values
of $\alpha$ and $d_0$. After each jump of $c$, its value remained
constant until the order parameter relaxed to the corresponding new
metastable state.} As seen from Figs.~\ref{clusters1}(b) to (d),
despite of a gradual compression of the cluster with increasing $c$,
its shape is preserved even at $c$ as large as 3.1. {Similar cluster
configuration and the same trends have been obtained, starting with
random distributions of a relatively large number of vortex pairs,
for different fixed values of $c$ ($c=0.7$, 1, 1.5, 2) and the same
set of other relevant parameters as well as for $c=1$ and 1.5 and a
two times larger sample ($L_x=L_y=200$).} At larger $c$, some of the
vortex pairs recombine, and smaller vortex clusters are stabilized
[see Figs.~\ref{clusters1}(e) and (f)] until the vortex-free state
is established at $c\approx 4.5$. {Besides a gradual compression of
the vortex cluster, increasing inward currents from the boundaries
lead also to the appearance of bulging vertical and horizontal
stripes in the density of the condensate.} One may notice that the
vortex-antivortex configuration in Fig.~\ref{clusters1}(f) resembles
that in Fig.~\ref{oscillating}(d). However, the former is fully
static: at $c=3.6$, shape-induced anisotropy of the confining
currents and the resulting anisotropy in the particle-density
distribution are sufficiently strong to prevent any changes in the
orientation of the ``vortex molecule''.

\begin{figure} \centering
\includegraphics*[width=1.0\linewidth]{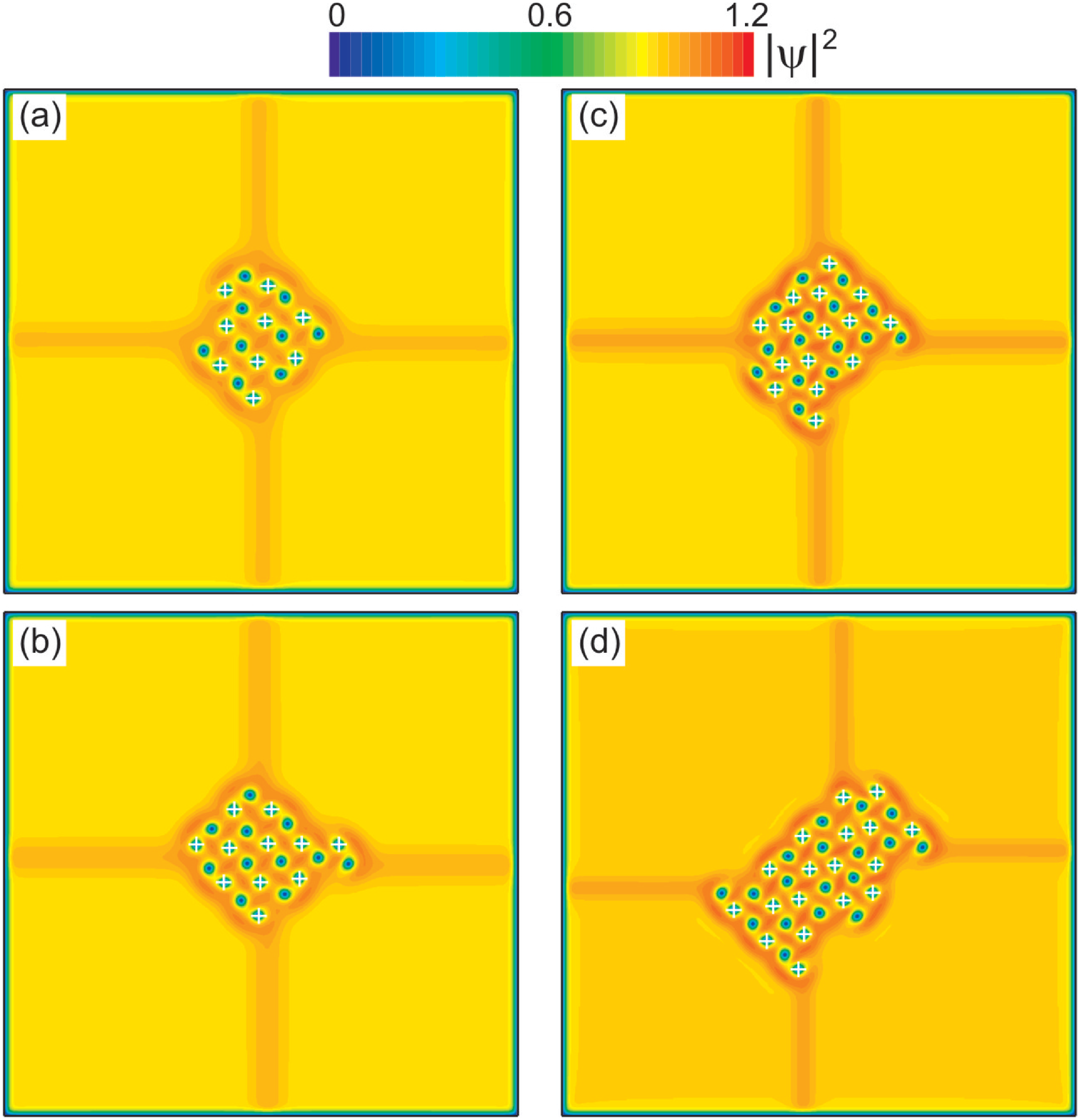}
\caption{Clusters with different number of vortex pairs $N$ in a
square with sides $L_x=L_y=200$ at $\nu=1$, $\kappa=0.05$ and
$\alpha=0.25$: $N=9$ for $c=1$, $d_0=0.6$ (a), $N=11$ for $c=1.2$,
$d_0=0.8$ (b), $N=15$ for $c=1.4$, $d_0=0.9$ (c), $N=18$ for
$c=1.5$, $d_0=1$ (d). The calculations are performed with the
Dirichlet boundary conditions, pumping intensity described by
Eq.~(\ref{pump}), and grid step $h=0.5$. The white crosses indicate
a vortex with counterclockwise circulating currents.
\label{clusters2}}
\end{figure}

As implied by Figs.~\ref{clusters1}(b) to (f), when increasing the
nonequilibrium parameter $c$ alone, the compressive effect, exerted
on a multivortex system by the inward currents from the sample
boundaries, increases more than the vortex-(anti)vortex repulsion,
caused by the radial currents from vortex cores. As a result, the
number of vortex pairs in the vortex cluster gradually decreases.
This does not mean, however, that strong deviations from equilibrium
inevitably impede formation of metastable vortex-antivortex states
with large numbers of vortex pairs. Indeed, excessively strong
currents from the sample boundaries can be reduced, e.g., by
increasing the width $d_0$ of the periphery regions with partially
suppressed pumping intensity [see Eq.~(\ref{pump})]. In
Fig.~\ref{clusters2} we show the metastable vortex configurations
obtained in the corresponding simulations. As further seen from this
figure, the number $N$ of vortex pairs in the metastable states,
formed as a result of vortex relaxation, tends to increase with
increasing $c$, provided that the inward currents from the sample
boundaries are appropriately tuned. This suggests that relatively
large ``vortex-antivortex lattices'' (possibly distorted due to
inhomogeneity and shape-induced anisotropy of confining currents)
can be stabilized in strongly nonequilibrium polariton condensates.

\section{Vortex-antivortex pair generation}

\begin{figure} \centering
\includegraphics*[width=0.9\linewidth]{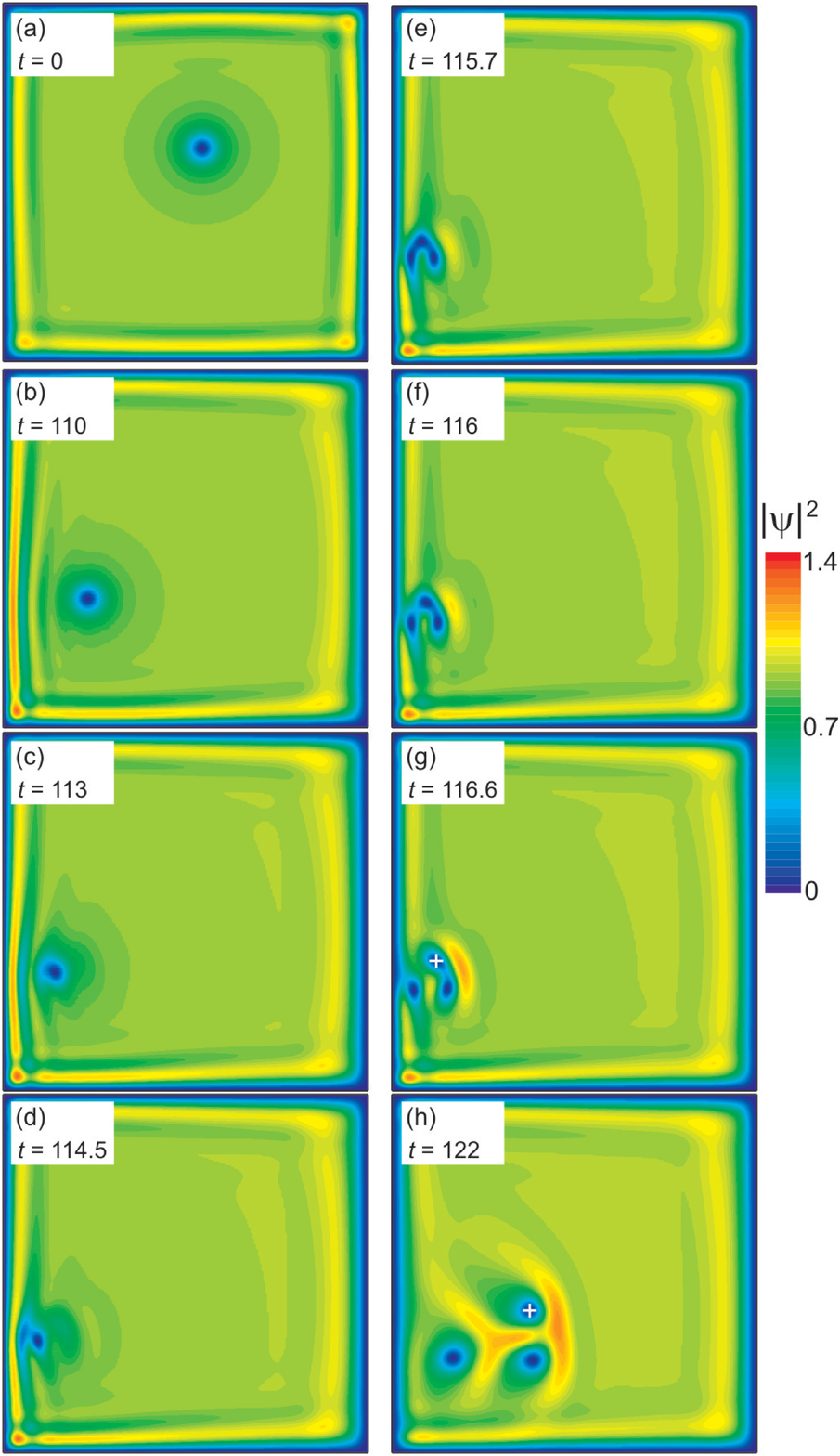}
\caption{Snapshots of the density distribution in a square with
sides $L_x=L_y=40$ at $c=4$, $\nu=1$, $\kappa=0.01$, $\alpha=0.4$,
and $d_0=1.3$ for different time $t$ after depinning of a vortex.
The white crosses indicate vortices with counterclockwise
circulating currents. The calculations are performed with the
Dirichlet boundary conditions, pumping intensity described by
Eq.~(\ref{pump}), and grid step $h=0.2$.  \label{gen1}}
\end{figure}

Let us turn to the last two questions raised in Sec. I: what happens
with self-accelerated vortices when further increasing the deviation
from equilibrium and can they produce new vortex-antivortex pairs?
We start with the case of a single vortex at a rather large value of
the nonequilibrium parameter, $c=4$. Initially, the vortex is pinned
to a random position in a finite-size square-shaped sample [see
Fig.~\ref{gen1}(a)]. At $t=0$ the pinning potential is removed and,
due to the combined effect of inward currents for the sample
boundaries and self-acceleration~\cite{30}, the vortex starts to
move towards the left-hand-side edge of the sample
[Figs.\ref{gen1}(b) and (c)]. When approaching the edge, the
relative velocity of the vortex core with respect to the inward
particle flow from the boundary strongly increases, the vortex
deforms [Figs.\ref{gen1}(c) and (d)], and eventually a new vortex
pair is nucleated [Figs.\ref{gen1}(e) and (f)] and spatially
separated due to vortex-(anti)vortex repulsion [Fig.\ref{gen1}(g)].
Then all the three topological defects are dragged away from the
edge by the confining current flow [Fig.\ref{gen1}(h)].

Qualitatively, the aforedescribed process may resemble vortex pair
generation induced by a fast moving obstacle in
superfluids~\cite{24,25,26} or by a defect surrounded by a
supercurrent in superconductors~\cite{29,28}. In the present case,
however, the ``defect'' itself produces circulating and radial
currents. This makes the process of new pair formation a bit more
complicated. Due to an interplay of different currents, an
additional suppression of the particle density develops in front of
the moving vortex core [see Fig.\ref{gen1}(d)]. In order to bypass
the resulting ''defect'' of a complex shape, the particle flow is
redistributed and the corresponding current crowding at the edges of
the defect ultimately leads to the nucleation of a new vortex pair.

\begin{figure} \centering
\includegraphics*[width=1.0\linewidth]{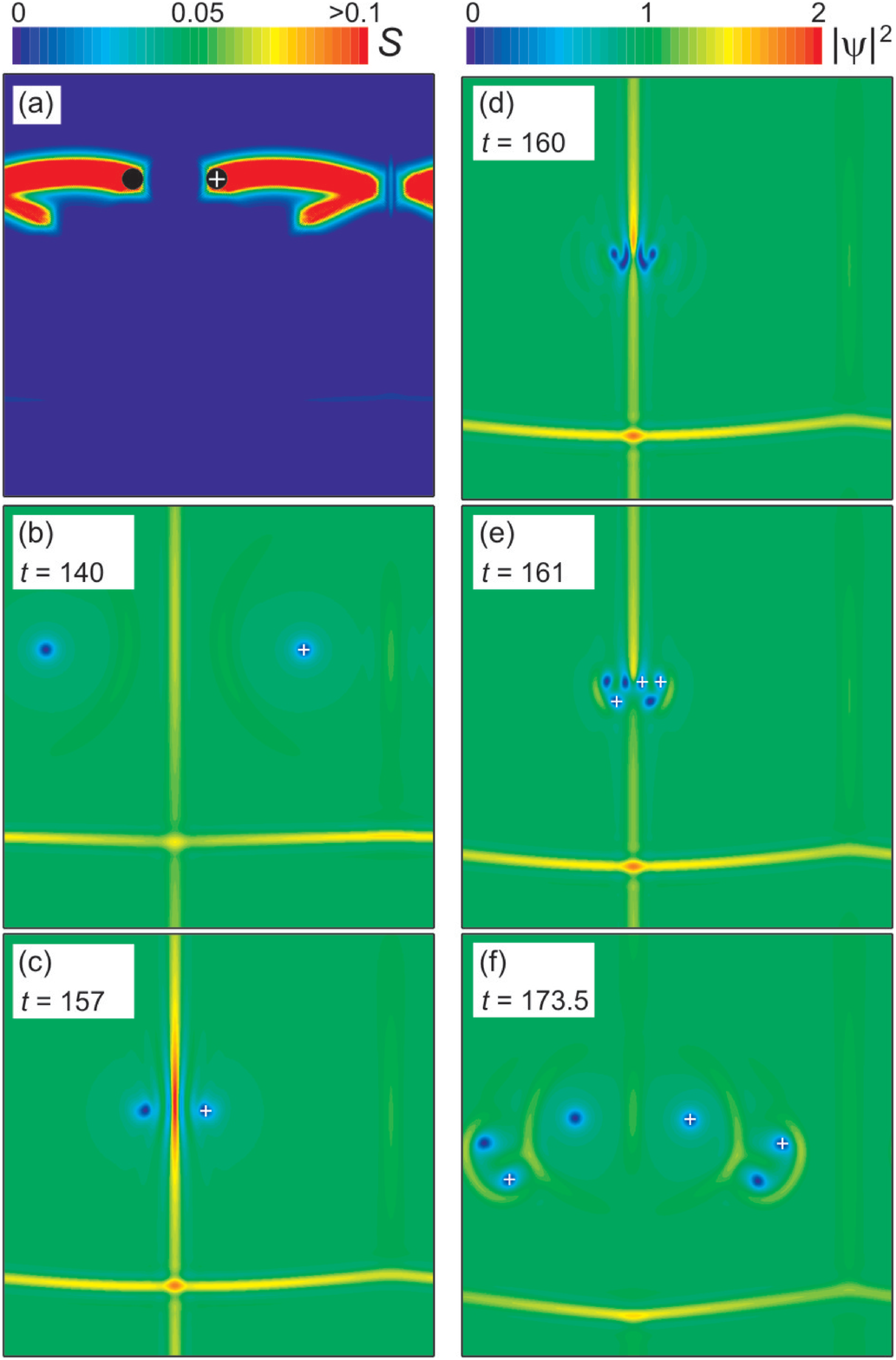}
\caption{Vortex dynamics in a unit cell of a periodic system with
period $L_x=L_y=80$ at $c=4$, $\nu=1$, and $\kappa=0.01$. (a)
Parameter $S$ for the time interval from the moment of depinning of
a vortex-antivortex pair ($t_1=0$) to $t_2=140$. The initial
positions of the vortex and antivortex are indicated with the black
circles. (b)-(f) Snapshots of the density distribution for different
time $t$ after depinning. The white crosses indicate vortices with
counterclockwise circulating currents. The calculations are
performed with the uniform pumping intensity $p\equiv 1$ and grid
step $h=0.2$.  \label{gen2}}
\end{figure}

Collision with the sample boundary is not the only possible source
of vortex pair generation by fast moving vortices. In
Fig.~\ref{gen2} we present the results of simulations for a
(boundaryless) periodic system, which initially contains one pinned
vortex-antivortex pair in a unit cell. After depinning, the vortex
and antivortex move along the trajectories shown in
Fig.~\ref{gen2}(a). Further dynamics of the system is represented by
few snapshots of the particle density distribution in
Figs.~\ref{gen2}(b) to (f). As seen from Fig.~\ref{gen2}(d), the
vortex-antivortex collision is accompanied by the formation of two
regions with strongly suppressed density $|\psi|^2$ in-between the
vortex and antivortex cores [cp. Fig.~\ref{gen1}(d)]. Similarly to
the case of the vortex-boundary collision, these suppressions
rapidly evolve into new vortex-anivortex pairs [see
Figs.~\ref{gen2}(e) and (f)].

\begin{figure} \centering
\includegraphics*[width=1.0\linewidth]{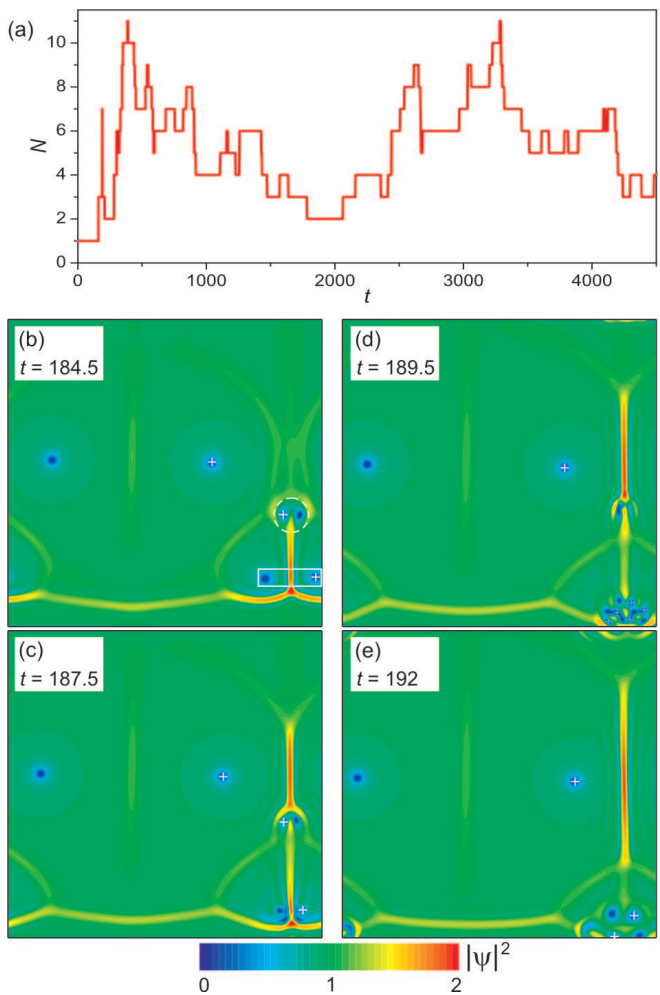}
\caption{(a) Number of vortex pairs per unit cell as a function of
time in a periodic system with period $L_x=L_y=80$ at $c=4$,
$\nu=1$, and $\kappa=0.01$. (b)-(e) Snapshots of the density
distribution for different time $t$ after depinning of the initially
present vortex-antivortex pair. The white crosses indicate vortices
with counterclockwise circulating currents. The vortex and
antivortex, shown inside the white dotted circle in panel (a),
further annihilate as a result of their collision. The collision
between the vortex and antivortex, shown inside the white solid-line
rectangle in panel (a), leads to production of new vortex pairs. The
calculations are performed with the uniform pumping intensity
$p\equiv 1$ and grid step $h=0.2$. \label{gen3}}
\end{figure}

In Fig.~\ref{gen3}(a), we show the dynamics of the number of vortex
pairs in the system under consideration on a relatively long time
scale. As implied by the fluctuating behavior of $N(t)$ in
Fig.~\ref{gen3}(a), vortex-antivortex collisions do not necessarily
lead to the generation of new vortex pairs. Very often, the result
of those collisions is more ``traditional'': the colliding vortex
and antivortex simply annihilate.

A more detailed analysis of the simulated vortex dynamics shows that
the generation of a vortex pair is more likely for higher total
current between the vortex and the antivortex. Sufficiently high
currents are reached only when the mean velocity of the vortex cores
is in the same direction as the circulating currents between vortex
and antivortex. For this reason, vortex-vortex collisions, where the
circulating currents between them are in opposite directions, do not
produce additional vortices.

{The condition for vortex pair generation} is fulfilled, e.g., in
the situation displayed in Figs.~\ref{gen2}(b) to (d). Another
example of this kind is the collision of the vortex and antivortex,
which are shown inside the white rectangular frame in
Fig.~\ref{gen3}(b). Interestingly, four rather than two new vortex
pairs nucleate as a result of this collision [see
Fig.~\ref{gen3}(c)]. However, three of them recombine within a
relatively short time interval [Fig.~\ref{gen3}(d)]. For the
vortex-antivortex pair, shown inside the dashed circle in
Fig.~\ref{gen3}(b), the above condition is violated and this pair
annihilates [see Figs.~\ref{gen3}(c) to (e)]. A vortex-antivortex
collision violating the above condition is evidenced also by the
shape the vortex trajectories in Fig.~\ref{gen2}(a). In this case,
the vortex and antivortex are simply ``scattered'' by each other due
to mutual repulsion.

\begin{figure} \centering
\includegraphics*[width=1.0\linewidth]{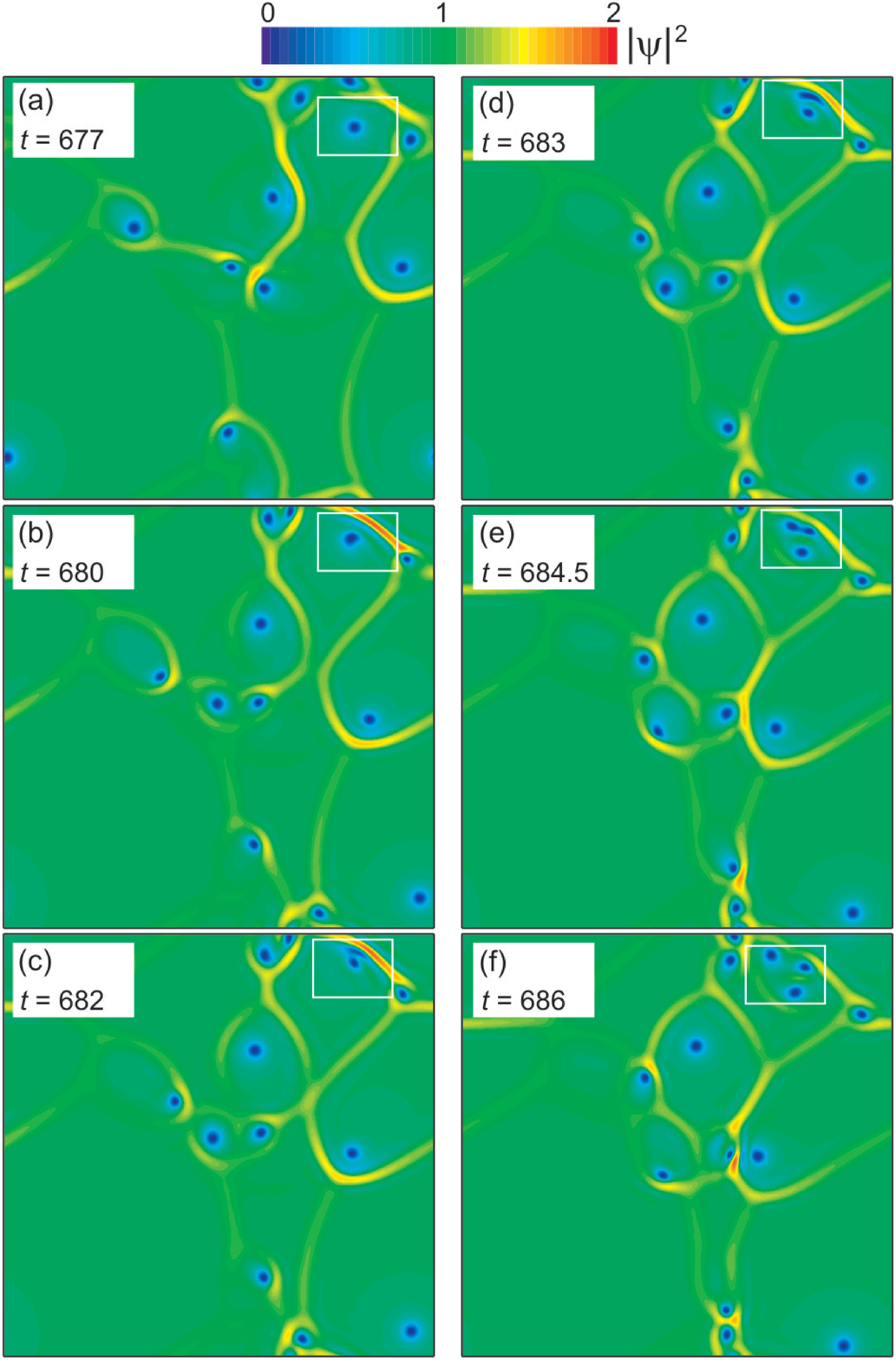}
\caption{Six snapshots of the density distribution for different
time $t$ after depinning of the initially present vortex-antivortex
pair in a periodic system with period $L_x=L_y=80$ at $c=4$,
$\nu=1$, and $\kappa=0.01$. The white solid-line rectangles indicate
the region where a fast moving vortex produces a new
vortex-antivortex pair. The calculations are performed with the
uniform pumping intensity $p\equiv 1$ and grid step $h=0.2$.
\label{gen4}}
\end{figure}

In a highly nonequilibrium polariton condensate, due to the presence
of strong radial flows, emitted from vortex cores, in combination
with a relatively slow relaxation of the flow fields~\cite{30},
vortex motion often lead to the appearance  of pronounced
time-dependent inhomogeneities in the current and density
distributions. An example can be seen in the formation of density
``bulges'' in front of fast moving vortices [see Figs.~\ref{gen1}(g)
and (h)]. Our simulations show that vortex pair generation can occur
also as a result of vortex interaction with those inhomogeneities.
This effect is illustrated in Fig.~\ref{gen4}. The vortex, which
moves in the region indicated with the white rectangle [see
Figs.~\ref{gen4}(a) and (b)], generates a new vortex pair when
approaching a density bulge [Figs.~\ref{gen4}(c) to (f)].
Remarkably, the growth of this bulge is seen to be caused, to a
great extent, by the fast motion of the same vortex [{compare}
Fig.~\ref{gen4}(b) to Fig.~\ref{gen4}(a)]. The shape of the curve
$N(t)$ in Fig.~\ref{gen3}(a), where the jumps with an increase of
$N$ by 1 predominate, implies that the aforedescribed mechanism of
vortex pair generation may be very efficient in multivortex systems.

It should be mentioned that in our simulations, performed for a
relatively wide range of parameters, we did not meet the situation
where a moving vortex would produce a fully developed
``vortex-antivortex street''. Still we do not completely exclude the
possibility that such a process can be realized under certain --
maybe rather exotic -- conditions.

\section{Conclusions}

We have numerically studied the dynamics of multivortex systems in
nonequilibrium quantum fluids described by the gGPE. It is shown
that, when moving away from equilibrium, the rate of
vortex-antivortex annihilation in these systems strongly decreases
due to an enhancement of the repulsive component in the
vortex-antivortex interaction.

In parallel with infinite periodic systems we have considered also
finite-size samples with the Dirichlet boundary conditions, where
vortex confinement is provided by the inward currents flowing from
the sample edges. In the case of weak damping, relaxation of
multivortex systems in those samples typically leads -- instead of
full annihilation of vortex pairs -- to the formation of metastable
vortex-antivortex clusters. The shape and size of these clusters are
determined by an interplay of the confining currents and
vortex-(anti)vortex repulsive interactions. Since the
vortex-antivortex repulsion is weaker than that between vortices of
the same chirality, relatively small and sparse clusters, obtained
at moderate deviations from equilibrium, often look to consist of
preformed ``bound'' vortex-antivortex pairs. The internal structure
of bigger and denser clusters, formed at larger values of the
nonequilibrium parameter, more resembles a vortex-antivortex
lattice, somewhat distorted due to (in general, anisotropic)
compressive effect of the confining currents.

We have also demonstrated that, at strong nonequilibrium,
self-accelerated vortices can induce generation of vortex pairs.
Thus, pair-production processes happen when a self-accelerated
vortex moves in a spatially inhomogeneous counterflow of particles.
Such a situation is realized, in particular, for vortices
approaching sample boundaries.  Another striking example is the
appearance of two or more new vortex pairs as a result of a
vortex-antivortex collision.

\section*{Acknowledgements}

We thank Sebastian Diehl for a stimulating discussion. This work was financially suppported by the FWO Odysseus program.

\end{document}